\documentstyle[12pt,cite]{article}
\catcode`@=11
\def\section{\@startsection {section}{1}{\z@}{-3.5ex plus -1ex minus 
 -.2ex}{2.3ex plus .2ex}{\large\bf}}
\def\subsection{\@startsection{subsection}{2}{\z@}{-3.25ex plus -1ex minus 
 -.2ex}{1.5ex plus .2ex}{\normalsize\bf}}
\catcode`@=12
\def\dalemb#1#2{{\vbox{\hrule height .#2pt
        \hbox{\vrule width.#2pt height#1pt \kern#1pt
                \vrule width.#2pt}
        \hrule height.#2pt}}}

\def\for{\lower6pt\hbox{$\Big|$}}

\begin{document}
\begin{flushright}
\hfill{DAMTP-R/52}\\
\end{flushright}
\renewcommand{\thefootnote}{\fnsymbol{footnote}}
\topmargin 0pt
\oddsidemargin 5mm
\begin{center}
{\Large\bf The Universality of M-branes\footnote[1]{Talk given at the
Imperial College Workshop on `Gauge Theories, Applied Supersymmetry and
Quantum Gravity', 5-10 July 1996.}}
\vspace{2.0truecm}

{\large G. Papadopoulos}

\vspace{1.0truecm}
DAMTP\\
Silver Street\\
University of Cambridge\\
Cambridge CB3 9EW.
\end{center}

\vspace{2.5truecm}

\begin{abstract}
     We review the evidence for the various dualities amongst the
five D=10 superstring theories and for the existence of M-theory using 
the associated
effective supergravity theories. We also summarise the combinatorial
 technics developed
for constructing BPS solutions in D=11 supergravity theory 
and conjecture that all the
BPS solutions of $D<11$ supergravity theories can be derived 
from the BPS solutions of
D=11 supergravity that preserve $1/2$ the supersymmetry.  
To demonstrate this, we derive
the dyonic p-brane solutions from eleven dimensions.
\end{abstract}
\vspace{.5truecm}

\newpage

\section{Introduction}

The last two years have seen remarkable progress towards understanding the
non-perturbative properties of superstring theory. Some of the main results
are the dualities found amongst all five D=10 superstring theories and the
evidence given for the existence of M-theory. The type IIA string
compactified
 on a
circle of radius $R_A$ is equivalent to the type IIB string~\cite{dhs,dlp}
compactified on a circle of radius $R_B$ provided that
$R_B=1/ R_A$. Similarly, the $E_8\times E_8$ heterotic
string theory compactified on a circle of radius $R_E$ and with gauge group
broken to
$SO(16)\times SO(16)$ with Wilson lines is equivalent to $SO(32)$ heterotic
string compactified on a circle of radius $R_{SO}$  with gauge group
again
 broken
to $SO(16)\times SO(16)$ with Wilson lines~\cite{nsw,nswa} provided that
$R_{SO}=1/R_E$. The transformation that establishes the equivalence in both
cases is T-duality which is a perturbative symmetry within the
superstring
 theory and
therefore it can be verified order by order perturbation theory.
Furthermore the $SO(32)$ heterotic string is S-dual to the the Type I
string~\cite{witten}, i.e. the behaviour of $SO(32)$ heterotic string at
strong string coupling is given by a weakly coupled type I string and 
vice-versa. x
Note though that equivalence of two theories under S-duality goes
beyond 
superstring
perturbation theory. 

One way to establish a relation between N=1 superstrings ($E_8\times
E_8$ 
and $SO(32)$
heterotic, and type I)  and N=2 superstrings (IIA and IIB) is
via M-theory; N denotes the number of supersymmetries. (For a recent review see
Schwarz~\cite{schw}.) There is no intrinsic definition of M-theory but
assuming that its
effective theory is the D=11 supergravity, it is possible to gather 
evidence which
support the conjecture that M-theory compactified on
$S^1$ is equivalent to the  IIA string~\cite{pkt,witten}; the compactification
radius is related to the IIA string coupling constant in such a way that the
strong coupling limit of IIA string is M-theory.  On the other hand, 
compactifying
M-theory on
$S^1/Z_2$ is equivalent to the
$E_8\times E_8$ heterotic string~\cite{hw}; similarly the strong coupling
limit of
$E_8\times E_8$ heterotic string is again M-theory. It is clear then
that all superstring theories can be derived from M-theory either as
Kaluza-Klein (KK) reductions or as KK reductions followed by T- and S-duality
transformations. 

Evidence for the existence of a duality symmetry between two
string theories can be found by examining the behaviour of the BPS
part 
of their
spectrum under the proposed duality. This is because only the BPS states are
expected to be stable under the changes in the coupling constants of the two
theories required by the duality. In the effective theory context, one
investigates the
classical solutions of the associated effective supergravity theories 
that preserve a
proportion of the supersymmetry; we shall call these solutions BPS 
solutions.  There
are many such solutions of $D<11$ supergravities.  However, since all 
superstring
theories in $D<10$ dimensions are compactifications of the $D=10$ ones
and the latter
are related to M-theory, all BPS solutions of the associated
supergravity 
theories
should have an M-theory interpretation.

We shall review the properties of the BPS solutions of the
supergravity 
theories
with emphasis in D=10 and D=11 dimensions and then we shall use them to explain
the following relations: (i) The KK interpretation of IIA string theory as
the reduction of  M-theory on $S^1$, (ii) the interpretation of  
$E_8\times E_8$
heterotic string as the compactification of  M-theory on $S^1/Z_2$,
(iii) the T-duality equivalence between type IIA and IIB strings, and finally
comment on the S-duality equivalence between the type I and SO(32) heterotic
strings. We shall then conjecture that all BPS solutions of $D<11$ supergravity
theories can be derived from the BPS solutions of  D=11 supergravity
theory 
that
preserve $1/2$ of the supersymmetry and we shall demonstrate this by deriving 
some of the dyonic p-branes in $D=2p+4$ from eleven dimensions. 
\section{p-brane combinatorics}

Some of the BPS solutions of a D-dimensional supergravity
theory (with maximal supersymmetry) are interpreted as p-branes, i.e.
they are p-dimensional objects lying within a D-dimensional spacetime
and 
are the
sources of a $(p+2)$-form field strength $F_{(p+2)}$.  For example the
0-branes are
particles that are the sources of the Maxwell field $F_2$, the
1-branes 
are strings
that are the sources of the 3-form field strength $F_3$ and so on. 
(For a review see
ref~\cite{gpb}.) The p-brane solutions of a supergravity theory
satisfy 
the killing
spinor equations preserving $1/2$ of the supersymmetry and their mass 
per unit volume
$M_p$ saturates the bound
\begin{equation}
 M_p\geq\beta  Q_p\ ,\label{bone}
\end{equation}
where 
\begin{equation}
 Q_p=\int_{S^{D-p-2}}\, \star F_{p+2}\ ,\label{btwo}
\end{equation} 
is the charge per unit volume and $\beta$ is a positive constant.  
A generic p-brane
solution of a D-dimensional supergravity theory is
\begin{eqnarray}
ds^2  &=& H^\gamma dx\cdot dx+H^\delta dy\cdot dy\nonumber\\
F_{(p+2) } &=& \epsilon_{p+1}\wedge dH^\rho\nonumber\\
\phi &=& H^\sigma \ ,\label{bthree}
\end{eqnarray}
where $x$ are the worldvolume coordinates of the p-brane spanning a
$(p+1)$-Minkowski spacetime, 
$y$ are the coordinates of the transverse space of the p-brane spanning  a
$(D-p-1)$-Euclidean space,
$H=H(y)$ is a harmonic function of the transverse space,
$\epsilon_{p+1}$ 
is the
volume form of the Minkowski spacetime,
$\phi$ are scalars and  $ \gamma, \delta, \rho, \sigma$ are real numbers.

Interpreting the BPS solutions of supergravity theories as extreme
p-branes 
has been
proved very successful for developing powerful combinatorial technics 
for constructing
new BPS solutions from the p-brane solutions (\ref{bthree}).
These involve the following: (i) The magnetic dual~\cite{tete,nepo} of
the $p$-brane in
(\ref{bthree}) is a  $\tilde p$-brane  which couples to
the  Poincar\'e dual of $F_{(p+2)}$ where $\tilde p=D-p-4$.
Dyonic p-branes are those for
which $p=\tilde p$ and therefore
$p=(D/2)-2$, {\sl i.e.} there are dyonic 0-branes (dyons) in D=4, dyonic
strings in D=6, dyonic membranes in D=8 and dyonic 3-branes in
D=10. In particular, since the Hodge star operator has real
eigenvalues 
in $D=6,10$
dimensions, the possibility arises for having self-dual strings and 
self-dual 3-branes
solutions. (ii) Any p-brane solution (\ref{bthree}) in D-dimensions
reduces to (p-k)-brane solution in $D-d$ dimensions  by wrapping the 
worldvolume
coordinates of the p-brane on the homology k-cycles of the
compactifying 
d-dimensional
space. For toroidal compactifications this can be done explicitly
since 
wrapping is just
the KK reduction of the solution along its worldvolume directions. (iii) Given 
$p_i$-branes,
$i=1,\dots,n$, solutions of a D-dimensional supergravity theory 
preserving $1/2$ of the supersymmetry,
a new solution can be constructed preserving $1/2^n$ of the 
supersymmetry which can be
interpreted as the intersection of these branes on a common k-brane 
denoted with
$(k|p_1,\dots,p_n)$. The existence of such solution is determined with certain
rules~\cite{gpa,strom} called `intersection rules' and the new
solution 
is constructed
by superposing the $p_i$-brane, $i=1,\dots,n$, solutions using 
the `harmonic function
rule'~\cite {tseytlin, jerome}. The latter rule involves the 
introduction of a harmonic
function
$H_i$ for each 
$p_i$-brane and the solution is determined by the observation that
if all but one of the harmonic functions is set to one, say $H_k$, 
the solution should
reduce that of
$p_k$-brane.  (iv) Given a
$p$-brane solution or a $(k|p_1,\dots,p_n)$ solution, we can find new 
solutions by
either boosting ($k\geq 1$) or superposing them with a KK monopole. 

\section{Applications to superstrings}

The p-brane solutions of D=11 supergravity are the
membrane~\cite{duff} 
and its magnetic
dual the fivebrane~\cite{guven} each preserving $1/2$ of the
supersymmetry. 
We shall
refer to them collectively as M-branes. Apart from the M-brane 
solutions of D=11
supergravity, there are many other BPS solutions that are 
intersections of M-branes. 
The M-brane intersection rules~\cite{gpa} are the following: 
1. two membranes can
intersect on a 0-brane and two fivebranes can intersect on a 3-brane, 
2. a membrane
and a fivebrane can intersect on a string and 3. two fivebranes can 
intersect on a
string. Apart from the intersecting M-brane solutions constructed using these
intersection rules there is another solution in D=11 that has the 
interpretation of a
membrane within a fivebrane, $(2|2,5)$, and preserves $1/2$ of the
supersymmetry~\cite{pktgp}; we shall investigate this solution 
in the next section 
in connection with the dyonic membranes.

In superstring theory, we can classify the various p-branes according to the
dependence of their mass per unit volume, $M_p$, from the string 
coupling constant
$\lambda_s$ as fundamental p-branes, Dirichlet p-branes or D-p-branes, and
solitonic
$p$-branes.   The only fundamental extended objects in string 
perturbation theory are
the strings with $M_1\sim 1$,
and the solitonic objects are their magnetic duals the 5-branes with
$M_5\sim \lambda_s^{-2}$.  An intermediate case is the D-p-branes for which
$M_p\sim\lambda_s^{-1}$. (For a review on D-branes from the string 
point of view see
Polchinski et al~\cite{polj} and for some novel
properties~\cite{wittenb}.) 
The D=10
superstring theories have the following p-brane solutions: (i) The 
heterotic string
contains a fundamental string and a solitonic 5-brane.  (ii) The 
type I string contain
a D-string and a D-5-brane. (iii) The type IIA string has a 
fundamental string and a
solitonic 5-brane as well as D-p-branes for
$p=0,2,6,8$ and (iv) the type IIB string has a fundamental string, a solitonic
5-brane and D-p-branes for $p=-1,1,3,5,7, 9$. (The D-p-branes for 
$p=-1$ and $p=9$ are
the gravitational instantons and the D=10 Minkowski spacetime, respectively.) 

The KK ansatz for compactifying D=11 supergravity on a circle to D=10 are
\begin{eqnarray}
ds_{(11)}^2  &=& e^{-{2\over3}\phi} ds_{(10)}^2+e^{{4\over3}\phi}
\big(dx_{11}+A\big)^2\nonumber\\
G &=& F_4+F_3\wedge dx_{11}\ ,\label{bfive}
\end{eqnarray}
where $G$ is the D=11 4-form field strength, the D=10  metric 
$ds_{(10)}^2$ is in the
string frame, $\phi$ is the dilaton,
$A$ is the KK-vector and the rest of the notation is self-explanatory.  Since
$\lambda_s=\exp<\phi>$, the radius of compactification is $R=\lambda_s^{2/3}$.
The M-theory interpretation of the IIA p-branes~\cite{pkt, witten} 
is as follows: the
IIA 0-branes are simply associated with the KK states of the 
compactification of
M-theory on
$S^1$, the IIA fundamental string is the wrapping of D=11 
membrane on $S^1$, the
IIA membrane is just the direct reduction of the D=11 membrane 
on $S^1$, the IIA
4-brane is the wrapping of the D=11 fivebrane on $S^1$, the IIA 
5-brane is the direct
reduction of the D=11 fivebrane on $S^1$ and the D=10 6-branes 
are the KK monopoles of
the compactification. (With the term `direct reduction' of a 
p-brane we mean the
wrapping of the p-brane on the 0-cycle of the compactifying space.)  
Finally the 
8-brane which is a solution of the massive IIA supergravity theory 
does not seem to
have a direct D=11 interpretation.  

As we have mentioned in the introduction, type IIA and type IIB strings are
equivalent under T-duality. T-duality acts on the various p-brane 
solutions of the
associated effective supergravity theories as follows: the 
fundamental IIB string and
solitonic IIB 5-brane transform under T-duality to the fundamental 
IIA string and to
the solitonic IIA 5-brane, respectively. The D-p-branes of either IIA or IIB
supergravity transform under T-duality to D-(p$\pm$1)-branes where 
the sign depends
on the choice of performing the T-duality transformation on a transverse or a
worldvolume directions of the D-p-brane, respectively. Thus, T-duality
transforms all
IIB p-brane solutions  to IIA p-brane solutions and vice-versa. This serves as
evidence  for the equivalence of type IIA and type IIB strings. Alternatively,
the equivalence of IIA and IIB can be used to derive the IIB p-branes
from the M-branes by first reducing the latter to the IIA p-branes
and then use the T-duality transformation to transform them to the IIB
p-branes.

So far we have used the various p-brane solutions preserving $1/2$ of
supersymmetry to provide evidence for M-theory/IIA
string and IIA/IIB strings dualities.  However the BPS solutions with
the interpretation of intersecting p-branes can also be used for the same
purpose~\cite{gpc}.  For example the $(0|0,1)$ and $(0|0,4)$ solutions in IIA
supergravity each preserving $1/4$ of supersymmetry have being
interpreted 
as KK 
modes on the string and on the 4-brane, respectively. (For 
the $(0|0,4)$ solution see
also ref~\cite{dkps}.)   Such solutions are expected because as we 
have seen, the IIA
string and IIA 4-brane are derived from the wrapping of D=11
membrane and fivebrane on $S^1$, and therefore from the IIA
perspective there must be KK modes on the string and the 4-brane.  

As in the case of IIA string, the strong coupling limit of 
$E_8\times E_8$ heterotic
string is also the M-theory~\cite{hw}. More precisely, 
M-theory compactified on the
orbifold 
$S^1/Z_2$ gives the $E_8\times E_8$ heterotic string where
$Z_2$ acts on the compactified coordinate $x_{11}$  as 
$x_{11}\rightarrow -x_{11}$ 
and on the 4-form field strength as $G\rightarrow -G$.  The compactified D=11
spacetime has as boundary the disjoint union of two D=10 
spacetimes one at each fixed
point of the
$Z_2$ action on $x_{11}$; the radius $R$ of compactification 
is proportional to the
distance between the two fixed points and $R=\lambda_s^{2/3}$.  
The D=11 membrane stretches between the two D=10 sheets of 
the boundary of the D=11
spacetime and intersects them at a string.  For small string 
coupling constant the two
D=10 spacetimes come close together and they become the D=10 
dimensional spacetime of
heterotic string.  On the other hand the D=11 fivebrane 
lies entirely within the
boundary of the D=11 spacetime and in the small string coupling constant limit 
becomes the heterotic 5-brane. The string and 5-brane are 
the only stable solutions
of the
$E_8\times E_8$ heterotic string effective theory. The rest of 
the p-branes that 
appear in the IIA theory are unstable in the heterotic case  
because the orbifold
symmetry
$Z_2$ projects out the associated form field strengths and therefore no such
supersymmetric configurations can exist. The $E_8\times E_8$ gauge group of the
heterotic string arises from an `anomaly' argument; we attach an 
$E_8$ gauge group to
each D=10 sheet of the boundary of D=11 spacetime so for weak 
string coupling we get
the $E_8\times E_8$ gauge group on the D=10 spacetime which is 
necessary for the
cancellation of anomalies of the heterotic string.   

Finally, it remains to examine the duality between the two 
heterotic strings, and the
$SO(32)$ heterotic string and the type I string.  In the first case T-duality
transforms the fundamental string and solitonic 5-brane of one theory to the
fundamental string and solitonic 5-brane of the other.  In the second
case the fundamental string and solitonic 5-brane of the $SO(32)$ heterotic
string under S-duality transform to the D-1-brane and D-5-brane of the type I
string and vice-versa~\cite{cmha, cmh}. 

\section{Dyonic p-branes from M-branes}

In the last two sections, we have derived some of the BPS solutions of
D=10 supergravities from the M-brane solutions of D=11 supergravity, 
we shall now 
turn to do the same for the dyonic p-branes of $D=(2p+4)$-supergravity
theories.  As
we have mentioned in the previous section the self-dual 
3-brane~\cite{lu} of IIB
theory can be derived by either the D=11 membrane or fivebrane.  
In the former case
we compactify the D=11 supergravity on $S^1$ and directly 
reduce the D=11 membrane to
the IIA 2-brane. Then we use T-duality to transform the IIA 2-brane to the IIB
self-dual 3-brane.  Alternatively, we also compactify D=11 
supergravity on $S^1$ but
this time we wrap the D=11 fivebrane on $S^1$ to the IIA 4-brane. 
Then again we use
T-duality to transform the IIA 4-brane to the IIB self-dual 3-brane.  
Next the dyonic
membrane solutions of D=8 N=2 supergravity can be obtained from M-theory in two
different ways.  First we start from the
$(2|2,5)$ solution of D=11 supergravity which has the interpretation 
of a membrane
within a fivebrane~\cite{pktgp}. We then compactify the D=11 supergravity on
$T^3$ wrapping $(2|2,5)$ along the three directions of the fivebrane 
orthogonal to
the membrane to get the dyonic membrane solutions of N=2 D=8 supergravity.
Alternatively, we compactify IIB on
$T^2$ and wrap the IIB self-dual 3-brane on the linear combination
\begin{equation}
 n \alpha_1+m \alpha_2\ ,\label{done}
\end{equation} 
of the fundamental homology cycles $\alpha_1, \alpha_2$ of $T^2$ 
where $(n,m)$ are
integers. It turns out that the self-dual 3-brane solution of IIB reduces the
dyonic membrane solution D=8 N=2 supergravity~\cite{hwa} with charge
$(n,m)$.
  Next, the self-dual string~\cite{lu} of D=6 N=4 supergravity 
can be obtained by
compactifying the IIB supergravity  on $K_3$ and wrapping the 
self-dual 3-brane on
the homology 2-cycles~\cite{wittenc} of
$K_3$.  Another way to derive the self dual string is to begin 
from the $(1|2,5)$
solution of D=11 supergravity with the harmonic function associated 
with the membrane
identified with the harmonic function associated with the fivebrane 
and perform the
following chain of KK reductions and T-duality transformations:
\begin{equation}
 (1|2,5)_M{\buildrel S^1\over\rightarrow} (1|2,4) {\buildrel T\over\rightarrow}
(1|3,3)_B{\buildrel T^4\over\rightarrow} 1^+ 
\ ,\label{dtwo}
\end{equation} 
where $(1|3,3)_B$ is a solutions of IIB supergravity and in the last 
step we have
wrapped the four worldvolume coordinates of the two 3-branes which 
are orthogonal to
the string on a linear combination of homology 2-cycles of $T^4$.  Finally the 
dyons of D=4 N=4 supergravity can be derived either by compactifying 
the D=8 N=2
supergravity on
$K_3$ and by wrapping the dyonic membrane solutions on the homology 
2-cycles of $K_3$
or by compactifying the D=6 N=4 supergravity on $T^2$ and by wrapping 
the self dual
string on the linear combination (\ref{dtwo}) of the fundamental 1-cycles of
2-torus~\cite{ver}.

\section{Summary}
The investigation of the strong coupling limit of type IIA and $E_8\times E_8$
heterotic strings has naturally led to M-theory which has as an
effective 
theory
the D=11 supergravity.  There is no intrinsic definition of M-theory but the
hypothesis that such theory exists  has been proved very useful to find a link
between the N=1 and N=2 D=10 superstrings and to develope a systematic way to
construct the BPS solutions of $D<11$ supergravity theories from the 
BPS solutions of
D=11 supergravity which preserve $1/2$ of the supersymmetry together 
with three rules
that determine their allowed intersections.

\bigskip

\newcommand{\Journal}[4]{{#1} {\bf #2}, #3 (#4)}
\newcommand{\NCA}{\em Nuovo Cimento}
\newcommand{\NIM}{\em Nucl. Instrum. Methods}
\newcommand{\NIMA}{{\em Nucl. Instrum. Methods} A}
\newcommand{\NPB}{{\em Nucl. Phys.} B}
\newcommand{\PLB}{{\em Phys. Lett.}  B}
\newcommand{\PRL}{\em Phys. Rev. Lett.}
\newcommand{\PRD}{{\em Phys. Rev.} D}
\newcommand{\ZPC}{{\em Z. Phys.} C}
\newcommand{\MPL}{{\em Mod. Phys. Lett.} A}

\end{document}